# Enhanced sputtering and incorporation of Mn in implanted GaAs and ZnO nanowires

A. Johannes[1], S. Noack[1], W. Paschoal Jr[2,3], S. Kumar[2,3], D. Jacobsson[2], H. Pettersson[2,3], L. Samuelson[2], K. A. Dick[4], G. Martinez-Criado[4], M. Burghammer[4] and C. Ronning[1]

[1]*Institute for Solid State Physics, Friedrich-Schiller-University Jena, Max-Wien-Platz 1, D-07743 Jena, Germany*

[2]*Solid State Physics/The Nanometer Structure Consortium, Lund University, Box 118, SE-221 00 Lund, Sweden*

[3]*Dept. of Mathematics, Physics and Electrical Engineering, Halmstad University, Box 823, SE-301 18, Halmstad, Sweden*

[4]*Centre for Analysis and Synthesis, Lund University, Box 124, S-221 00 Lund, Sweden*

[5]*European Synchrotron Radiation Facility, F-38043 Grenoble, France*

corresponding author: andreas.johannes@uni-jena.de

Abstract:

We simulated and experimentally investigated the sputter yield of ZnO and GaAs nanowires, which were implanted with energetic Mn ions at room temperature. The resulting thinning of the nanowires and the dopant concentration with increasing Mn ion fluency were measured by accurate scanning electron microscopy (SEM) and nano-X-Ray Fluorescence (nanoXRF) quantification, respectively. We observed a clear enhanced sputter yield for the irradiated nanowires compared to bulk, which is also corroborated by *iradina* simulations. These show a maximum if the ion range matches the nanowire diameter. As a consequence of the erosion thinning of the nanowire, the incorporation of the Mn dopants is also enhanced and increases non-linearly with increasing ion fluency.

**Introduction**

Semiconductors show a vast range of functionality, especially due to the possibility of changing material properties by doping. Ion beam implantation is a well established doping method at a scientific and industrial level. Implantation into nanostructures, however, only came into focus with the recent ascent of nanotechnology in general. Ion beam doping is an especially interesting option for nanowires (NWs), as doping during growth is challenging. Growth dynamics may prevent homogeneous dopant incorporation [1, 2] and the NW growth process itself is affected considerably by adding dopants to the growth environment [3, 4].

Ion beam implantation is free of some of these constraints as dopants are introduced after NW growth. Also chemical considerations and solubility only play a role in the actual activation and incorporation at desired sites within the semiconductor lattice, they do not principally impede the synthesis. Ion beam implantation does have the drawback that the impinging ion invariably introduces damage to the lattice. Initially this will take the form of interstitials and vacancies, but it may agglomerate to dislocation loops, defect clusters and eventually amorphization of the matrix [5, 6, 7]. Although NWs do show enhanced dynamic annealing [8], careful considerations of implantation parameters, in conjunction with modeling, are required to achieve homogeneous doping and minimal damage.

Various simulation programs exist for the simulation of ion irradiation of bulk materials, most notably *SRIM* is a popular implementation of the TRIM code [9]. Further expansions include the dynamic mixing and compositional changes during the implantation process [10] for bulk and thin film materials. However, new effects become very important if the structure size matches the ion range, which is the case for nanomaterials. The correct dopant and damage distribution can only be explained with software that considers the 3D nanostructure of the implanted sample. The recently developed, open source software *iradina* [11] can efficiently simulate the typical implantation situation in NWs. A recent further development considering the implantation dynamics of 3D nanostructures is the Tri3Dyn code [12].

This work investigates the enhanced sputter yield of NWs, predicted by the Monte Carlo simulation software *iradina* for implantation into nanostructures, by measuring the change in radius of GaAs NWs by ion beam implantation of Mn with a SEM. We also evaluate the dopant concentration with accurate nano-X-Ray Fluorescence (nanoXRF) quantification of Mn implanted ZnO NWs. These material combinations were chosen as they are both of interest as potential diluted magnetic semiconductors.

**Simulations**

The simulation of the implantation conditions was performed using *iradina* [11] for a GaAs NW of 90 nm diameter. The x and y discretization was 0.5 nm, while in z direction, parallel to the NW axis, the voxels were 10 nm long and periodic boundary conditions were set. The simulation results do not vary significantly with different discretizations.

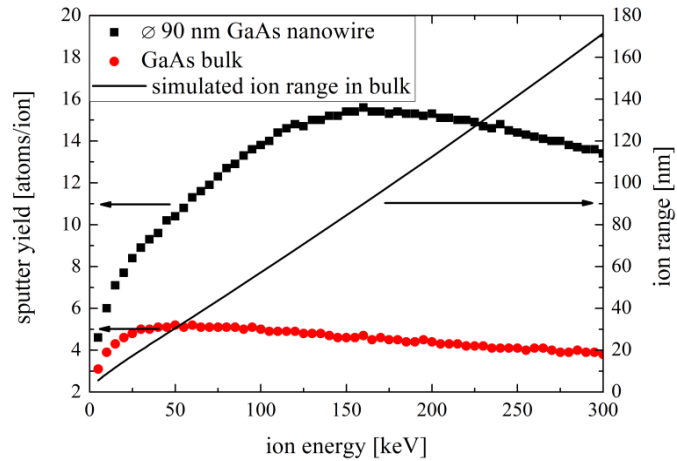

Figure 1:  Sputter yield and ion range of Mn ions implanted at an angle of 45° in GaAs as a function of the ion energy. The black squares denote the values for a NW with 90 nm diameter, while the red circles show the bulk value.

In figure 1 the simulated sputter yield is plotted for various ion energies. Here, the black squares show the results from implantation at 45° angle to the NW axis, while the red circles show the sputter yield for a bulk sample simulated with the same parameters. The solid black line shows the ion range within bulk according to SRIM [9], it reaches 90 nm at about 150 keV. Also, at this energy, where the ion range matches the NW diameter, the sputter yield reaches a clear maximum. This can be understood as the energy for which the collision cascade fills on average the whole NW. Much of the introduced energy is thus available for sputtering atoms at the surface, both in forward and in backward direction. For lower ion energies, the increased sputter yield compared to bulk is caused by an increased effective angle of incidence when implanting into the curved surface of a NW compared to the flat surface of bulk samples. No particles are sputtered forward. For larger ion energies the probability that the impinging ion can fly all the way through the wire increases. In these cases the ion will not deposit all its energy and sputtering is reduced.

From the same arguments it is clear that a maximum will also appear when the sputter yield is plotted versus the diameter of a NW for a fixed energy, instead of a varying ion energy and fixed diameter.

Such a situation can be much easier to realize experimentally as NWs of varying diameters can be implanted simultaneously, while different ion energies require many implantation runs. The respective simulation data for such a case are plotted later on together with the evaluated sputter yield results. Similar simulations were also done for the implantation of 175 keV Mn into 200 nm diameter ZnO NWs and resulted into analogous results.

**Experimental**

GaAs NWs were grown on ⟨111⟩ GaAs substrates by MOVPE from mono-disperse 80 nm Au catalyst particles. The samples were doped during growth with Zn, which is important for the electrical properties but not for this study, as it has no influence on the morphological changes investigated in this study. More details of the growth and properties of these samples are described elsewhere [13, 14]. The samples consist of a relatively sparse distribution of NWs with a length of 2 µm and a diameter of ~80 nm, as shown in the scanning electron microscopy (SEM) image in figure 2(b).

Additionally, ZnO NWs were grown on Si wafers with a layer of ~450 nm sputtered aluminium doped ZnO (AZO) via vapour transport [15]. The AZO layer grows preferentially in c-direction and thus the subsequent wires grow aligned in a self-catalytic epitaxial process on top of this substrate. These samples show a dense and aligned array of > 10 µm long and ~ 200 nm thick wires, as shown in figure 2(c).

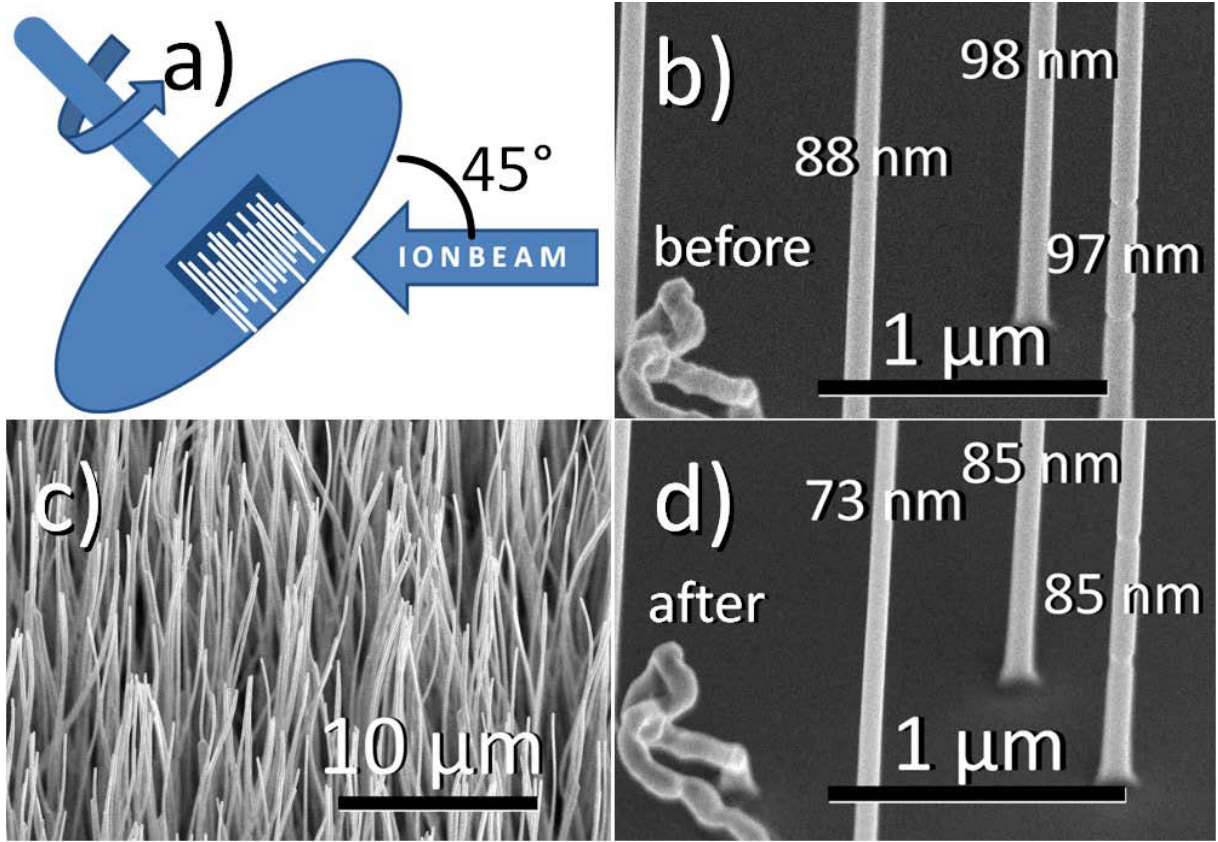

Figure 2: (a) Schematic view of the implantation scheme where the sample is rotated under the ion beam at an angle of 45°. A SEM image of GaAs wires (b) before and (d) after implantation of $1.3 \times 10^{16}$ Mn$^+$ ions/cm$^2$ at 40 keV. The diameters were measured with a high resolution SEM on exactly the same wires before and after the implantation process. (c) SEM image of ZnO NWs after implantation of $2.4 \times 10^{16}$ Mn$^+$ ions ions/cm$^2$ at 175 keV.

Figure 2 (a) shows the implantation setup characterized by a 45° tilted sample, which is rotated under the ion beam. These implantation conditions ensure homogeneous doping and avoids wire bending, which would happen if the wires were implanted from one side only [16]. Figure 2 (b) shows that the GaAs NW remain straight after ion implantation of $1.325 \times 10^{16}$ Mn$^+$ ions/cm$^2$ at 40 keV and 300°C. The elevated temperature ensures that the NWs remain crystalline [8]. According to the *iradina* simulation results this corresponds to a Mn concentration of about 2.3 at %. The experimental sputter yield (SY) is calculated according to

$$SY = \frac{\Delta V \cdot \rho_{at}}{A \cdot N_i} = \frac{2\pi(R^2 - r^2) \cdot \rho_{at}}{(R-r) \cdot N_i}, \quad (1)$$

where ΔV is the reduction of the volume the of the wire, $\rho_{at}$ the atomic density, A the irradiated area, R and r the radii before and after irradiation and $N_i$ the ion fluency. Sputtering from the top facet of the wires is neglected.

The mean diameter of the ZnO NWs is ~ 200 nm ± 50 nm, lengths are > 10 µm and the sample coverage is quite dense as can be seen in the SEM image figure 2 (c) of a sample after irradiation. The simulation yields in this case a homogeneous doping profile for $Mn^+$ ions implanted at an ion energy of 175 keV. Samples were irradiated with $2.38 \times 10^{14}$, $2.38 \times 10^{15}$, $2.38 \times 10^{16}$, $4.76 \times 10^{16}$, $9.53 \times 10^{16}$ and $1.91 \times 10^{17}$ ions/cm$^2$ to yield concentrations of 0.01, 0.1, 1, 2, 4 and 8 at % Mn in ZnO, respectively, according to the *iradina* simulations. This implantation was done at room temperature, as ZnO remains crystalline at these implantation conditions. The NWs were imprinted onto copper mesh, carbon foil TEM grids for single NW inspection by both SEM and nanoXRF.

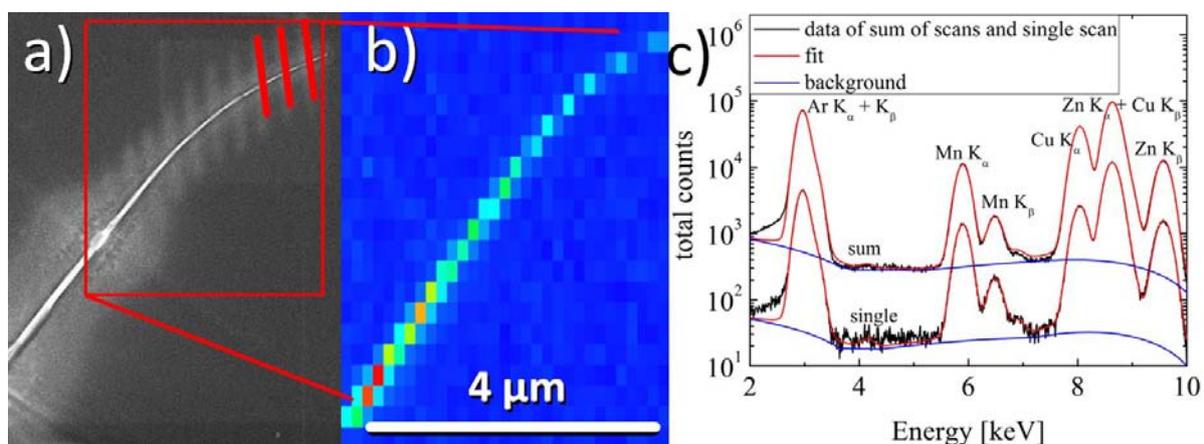

Figure 3: (a) SEM image and (b) corresponding XRF map of a single Mn-implanted ZnO NWs. The red lines indicate where longer exposure scans were performed, and where marginal modifications induced by the high intensity synchrotron beam are visible. The wire remains intact except for a spot in the lower left where the beam exposure was prolonged during the setup of the measurement protocol. (c) PyMCA fits of the obtained XRF data from the scans indicated in (a). The top line shows fitting to the sum of all scans, while the bottom line shows the integrated spectrum of a single scan.

At the European Synchrotron Radiation Facility (ESRF) ID13 micro focus beam line, a 14.9 keV focused beam with a diameter of 250-300 nm was scanned across single NWs. A Vortex EM silicon drift X-ray detector was placed near the sample to collect the emitted fluorescence X-rays (XRF) as a function of position. Figure 3 (a) shows an SEM image of an investigated BW after the XRF scans. This particular NW shows only some slight indications of damage by the intense synchrotron beam. All other investigated NWs in this study showed less indication of damage. Figure 3 (b) shows a map of the X-ray counts used to identify a single NW. Scans across the NW diameter were made at multiple points along the wire length, as indicated by the red lines in figure 3(a). With a step size of 50 nm and a 10 s integration time per step, average counts of more than $10^5$ per scan were obtained to allow accurate quantification. A few scans were rejected as they contained fewer counts. They were deemed to have missed the wire due to sample drift or misalignment. Figure 3 (c) shows an example of an obtained XRF spectrum of such a scan. The background subtraction, fitting and quantification were done with the PyMCA software package [17].

**Results**

The evaluation of the diameters taken from the SEM analysis of GaAs NWs before and after Mn implantation are shown in figure 4 (a). Around 60 individual NWs were investigated before and after the implantation of $1.325 \times 10^{16}$ ions/cm$^2$ Mn$^+$ ions at 40 keV and 300°C . The NWs show an average diameter of 82 ± 14 nm before and 66 ± 14 nm after implantation. From the individual differences, the reduction in diameter can be evaluated to be 16.5 ± 3.4 nm. The accuracy of individual measurements of the diameter with SEM can be estimated to be around 2 nm. This uncertainty contributes to the spread in calculated sputter yields plotted in figure 4 (b) as a function of the diameter of the NWs. The sputter yield was calculated according to equation 1. Averaging the sputter yields over all diameters gives a value of 8.7 ± 1.8. The error approximation is the standard deviation in each case. This value is significantly higher than the simulated bulk value of 5.1 plotted in figure 4 (b) as a red line and confirms the enhanced sputtering of NWs compared to bulk and thin film systems during ion implantation. The red circles mark the sputter yield obtained from *iradina* simulations. The pronounced maximum at ~ 25 nm corresponds to the ion range of 40 keV Mn in GaAs, as discussed previously. The measured data is indeed inclined toward lower sputter yields for larger diameters; however, a clear confident reproduction of the simulated curve is not possible with these results.

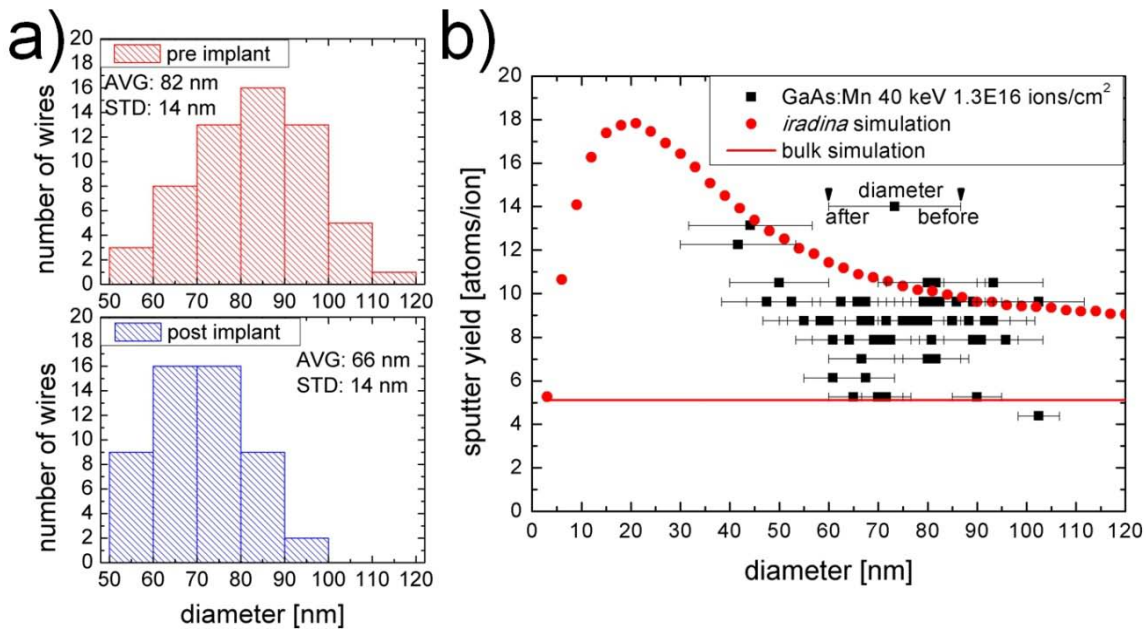

**Figure 4:** (a) Histogram of the SEM measured NW diameters before and after the implantation of Mn$^+$ ions with an ion energy of 40 keV. In (b) the individual measured diameters are plotted together with the resulting sputter yield, as function of the diameter. The beginning and end of the 'error bar' mark the before and after implantation diameter. The red dots show the *iradina* simulated sputter yield for a NW, while the red line is the bulk sputter value under 45° incidence.

Figure 5(a) shows the XRF quantification of the Mn content in the implanted ZnO NWs as a function of ion fluence. The red line shows the extrapolation from the *iradina* simulation, as one would expect: the concentration increases linearly with the ion fluence. The black circles show the result from fitting the data of the sum of all scans, which however do not follow the expected trend. Furthermore, figure 5 (b) shows that there is a clear concentration profile along the length of the NWs, which was also not expected. The highest concentration was always found at the tip of the NWs and the lowest at the point where the NWs broke of the substrate. The highest concentration values are also plotted in Figure 5 (a) as blue triangles. It can be clearly seen that the concentration increases faster with increasing fluency than expected from the *iradina* simulation. This effect is even more pronounced at the tip of the NWs than averaging over the whole wire length.

The increased concentration with respect to the simulation can be understood by the strong and enhanced sputtering from the NW during the implantation process, reducing the wire diameter successively and thus eroding the host matrix material while adding dopants. This leads to a non-linear increase in concentration with fluency. Note that the increase is not strictly exponential, as also Mn is sputtered from the wire, so that the concentration can theoretically saturate. Thus, also the diameters are drastically reduced for fluencies larger than $1 \times 10^{16}$ ions/cm$^2$.

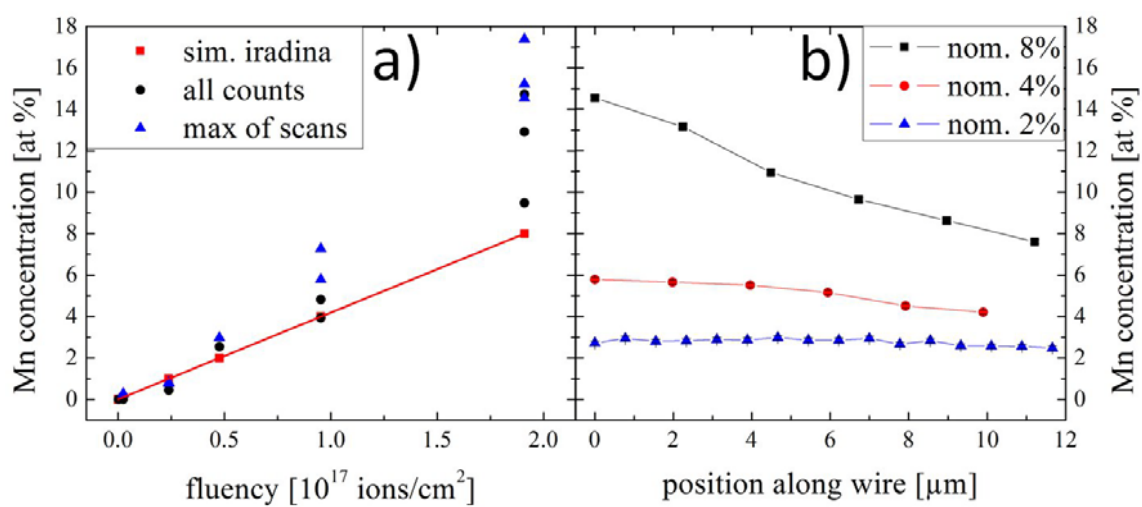

Figure 5: (a) Plot of the Mn concentration determined by XRF versus total implanted fluency for Mn-implanted ZnO NWs. The red line shows the result expected from the linear extrapolation of the *iradina* simulation. The black circles mark the concentration determined by fitting all XRF counts, while the blue triangles show only the scan with the highest concentration. (b) Representative Mn concentration along individual NWs plotted versus position for 3 different nominal concentrations.

The unexpected concentration profile along the wire length can also be attributed to the high density of the NWs on the sample during ion implantation, as shown in figure 2(c). Further down a given NW, towards the substrate, the probability rises that the wire is just shadowed from the ion beam by one of its neighbours. An indication that this is the case is also the partial bending of the wires seen in figure 1 (c), as unequal shadowing from different sides will cause the wires to bend under ion irradiation. In addition to the shadowing, also material sputtered from neighbouring wires can be re-deposited on the NW surface. This is more pronounced at the bottom part of the NW than at its tip, as here the NW surface is exposed to more neighbouring NWs emitting sputtered atoms. In agreement with these two effects, the SEM image in figure 3 (a) shows that the NWs are clearly tapered after the implantation process, whereas they are straight prior implantation (not shown). This is attributed to a decreased effective ion fluency by shadowing and increased re-deposition of sputtered material from other wires further down the NW, than at its tip. Both these effects lead to a variation of the Mn concentration along the wire roughly by a factor of 2, as shown in figure 5 (b). The simulation cannot consider either of these effects, the data from the tips of the wires has to be considered the best for the comparison between the experimental results and the simulation. However, this finding is an interesting result in itself as it shows that significant and large effects arise from the sample morphology also on a larger scale than the single wire. It further demonstrates the necessity to consider the distances of NWs during ion implantation, and more important the use of dynamic simulation tools for the ion implantation of nanomaterials, such as the Tri3Dyn code [12].

**Conclusions**

We have shown that although ion beam implantation is a well established tool in semiconductor industry, physics, and technology, care needs to be taken when considering the ion beam interaction with nanostructures. The SEM investigation of single NWs before and after ion irradiation shows greatly enhanced sputtering compared to bulk or thin film systems. This can be understood by a Monte-Carlo simulation implemented in the *iradina* program as an enhanced interaction of the collision cascade of the impinging ion with the NW surface leading to both forward and backward sputtering. The maximum sputter yield is reached were the ion range is comparable to the NW diameter. During the implantation, the volume of the NWs is greatly reduced by enhanced sputtering, while dopants are continuously incorporated. In this way the enhanced sputtering in nanostructures leads to a non-linear increase in dopant concentration with ion fluency as investigated by nanoXRF. Thus, with enhanced sputtering in nanostructures, the need for dynamic simulation i.e. the need to include sputtering in the simulation to accurately predict dopant concentrations arises at lower fluencies than in bulk.

**Acknowledgements**

The authors gratefully acknowledge funding by the German Research Society (DFG) within the DACH project. We thank the European Synchrotron Radiation Facility (ESRF) for the allocated beamtime. The authors acknowledge financial support from the Nanometer Structure Consortium at Lund University (nmC@LU), the Swedish Research Council (VR), and the Knut and Alice Wallenberg Foundation (KAW).

[1]     Perea D E, Wijaya E, Lensch-Falk J L, Hemesath E R and Lauhon L J 2008 *Journal of Solid State Chemistry* **181** 1642–1649 ISSN 0022-4596 WOS:000258562900019

[2]     Connell J G, Yoon K, Perea D E, Schwalbach E J, Voorhees P W and Lauhon L J 2013 *Nano Letters* **13** 199–206 ISSN 1530-6984 WOS:000313142300035

[3]     Pan L, Lew K K, Redwing J M and Dickey E C 2005 *Journal of Crystal Growth* **277** 428–436 ISSN 0022-0248 http://www.sciencedirect.com/science/article/pii/S0022024805001259

[4]     Borgström M T, Norberg E, Wickert P, Nilsson H A, Trägårdh J, Dick K A, Statkute G, Ramvall P, Deppert K and Samuelson L 2008 *Nanotechnology* **19** 445602 ISSN 0957-4484 http://-iopscience.iop.org/0957-4484/19/44/445602

[5]     Wesch W 1992 *Nuclear Instruments and Methods in Physics Research Section B: Beam Interactions with Materials and Atoms* **68** 342–354 ISSN 0168583X http://linkinghub.elsevier.com/-retrieve/pii/0168583X92961058

[6]     Wang S X, Wang L M and Ewing R C 2001 *Physical Review B* **63** 024105 ISSN 0163-1829 WOS:000166382200016

[7]     Ronning C, Borschel C, Geburt S and Niepelt R 2010 *Materials Science & Engineering R-Reports* **70** 30–43 ISSN 0927-796X WOS:000285706100001

[8]     Borschel C, Spindler S, Lerose D, Bochmann A, Christiansen S H, Nietzsche S, Oertel M and Ronning C 2011 *Nanotechnology* **22** 185307 ISSN 0957-4484 WOS:000288653300010

[9]     Ziegler J F, Ziegler M D and Biersack J P 2010 *Nuclear Instruments & Methods in Physics Research Section B-Beam Interactions with Materials and Atoms* **268** 1818–1823 ISSN 0168-583X WOS:000278702300028

[10]     Möller W and Eckstein W 1984 *Nuclear Instruments and Methods in Physics Research Section B: Beam Interactions with Materials and Atoms* **2** 814–818 ISSN 0168-583X http://www.sciencedirect.com/science/article/pii/0168583X84903215

[11]     Borschel C and Ronning C 2011 *Nuclear Instruments & Methods in Physics Research Section B-Beam Interactions with Materials and Atoms* **269** 2133–2138 ISSN 0168-583X WOS:000294936400016

[12]     Möller W 2014 *Nuclear Instruments and Methods in Physics Research Section B: Beam Interactions with Materials and Atoms* **322** 23–33 ISSN 0168-583X http://www.sciencedirect.com/science/article/pii/S0168583X13011968

[13]     Paschoal W, Kumar S, Borschel C, Wu P, Canali C M, Ronning C, Samuelson L and Pettersson H 2012 *Nano Letters* **12** 4838–4842 ISSN 1530-6984 WOS:000308576000069

[14]     Kumar S, Paschoal W, Johannes A, Jacobsson D, Borschel C, Pertsova A, Wang C H, Wu M K, Canali C M, Ronning C, Samuelson L and Pettersson H 2013 *Nano Letters* **13** 5079–5084 ISSN 1530-6984 WOS:000327111700014

[15]     Borchers C, Müller S, Stichtenoth D, Schwen D and Ronning C 2006 *The Journal of Physical Chemistry B* **110** 1656–1660 ISSN 1520-6106 http://dx.doi.org/10.1021/jp054476m

[16]     Borschel C, Messing M E, Borgstrom M T, Paschoal W, Wallentin J, Kumar S, Mergenthaler K, Deppert K, Canali C M, Pettersson H, Samuelson L and Ronning C 2011 *Nano Letters* **11** 3935–3940 ISSN 1530-6984 WOS:000294790200073

[17]     Solé V A, Papillon E, Cotte M, Walter P and Susini J 2007 *Spectrochimica Acta Part B: Atomic Spectroscopy* **62** 63–68 ISSN 0584-8547 http://www.sciencedirect.com/science/article/pii/S0584854706003764

# Corrigendum: Enhanced sputtering and incorporation of Mn in implanted GaAs and ZnO nanowires (2014 J. Phys. D: Appl. Phys. 47 394003)

A. Johannes[1], S. Noack[1], W. Paschoal Jr[2,3], S. Kumar[2,3], D. Jacobsson[2], H. Pettersson[2,3], L. Samuelson[2], K. A. Dick[4], G. Martinez-Criado[4], M. Burghammer[4] and C. Ronning[1]

[1]*Institute for Solid State Physics, Friedrich-Schiller-University Jena, Max-Wien-Platz 1, D-07743 Jena, Germany*

[2]*Solid State Physics/The Nanometer Structure Consortium, Lund University,*

*Box 118, SE-221 00 Lund, Sweden*

[3]*Dept. of Mathematics, Physics and Electrical Engineering, Halmstad University,*

*Box 823, SE-301 18, Halmstad, Sweden*

[4]*Centre for Analysis and Synthesis, Lund University, Box 124, S-221 00 Lund, Sweden*

[5]*European Synchrotron Radiation Facility, F-38043 Grenoble, France*

corresponding author: andreas.johannes@uni-jena.de

In our recent publication [1] on the enhanced sputtering in nanowires we implemented an error in the MC simulation code *iradina* [2]. Due to this mistake the respective simulated sputter yields in figure 1 and 4 of ref. [1] are too low. We recalculated the sputter yields using the corrected version of the software yielding higher sputter yields shown in figure 1 and 2, to replace figures 1 and 4 of ref. [1]. Also the experimentally determined sputter yield must be corrected, as the used formula is correctly given by:

$$SY = \frac{\Delta V \cdot \rho_{at}}{A \cdot N_i} = \frac{2\pi(R^2 - r^2) \cdot \rho_{at}}{\sin(45°) \cdot (R+r) \cdot N_i}, \quad (1)$$

where $\Delta V$ is the reduction of the volume the of the wire, $\rho_{at}$ the atomic density, $A$ the irradiated area, $R$ and $r$ the radii before and after irradiation and $N_i$ the ion fluency as in ref. [1]. The correction on the irradiated area $A$ for the irradiation angle by $\sin(45°)$ was missing. This gives a higher average sputter yield of 12.3 ± 2.5. These changes do not affect the discussion in the text of ref. [1], as the relation between experiment and simulation remains similar. We would like to thank Prof. W. Möller for analysing our data and making us aware of the possibility of an error.

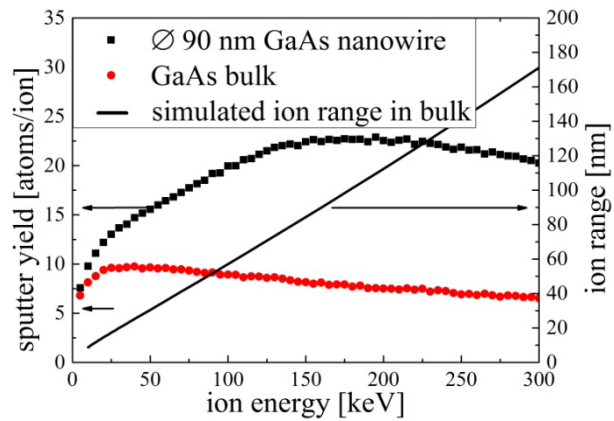

**Figure 1:** (Figure 1 in Ref. [1]) Sputter yield and ion range of Mn ions implanted at an angle of 45° in GaAs as a function of the ion energy. The black squares denote the values for a NW with 90 nm diameter, while the red circles show the bulk value.

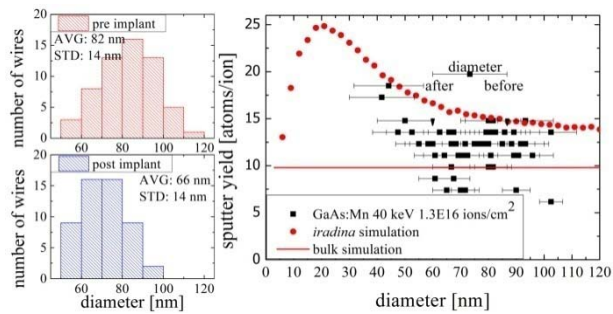

**Figure 2:** (Figure 4 in Ref. [1]) (a) Histogram of the SEM measured NW diameters before and after the implantation of Mn$^+$ ions with an ion energy of 40 keV. In (b) the individual measured diameters are plotted together with the resulting sputter yield, as function of the diameter. The beginning and end of the 'error bar' mark the before and after implantation diameter. The red dots show the *iradina* simulated sputter yield for a NW, while the red line is the bulk sputter value under 45° incidence.

[1]     A. Johannes, S. Noack, W. P. Jr, S. Kumar, D. Jacobsson, H. Pettersson, L. Samuelson, K. A. Dick, G. Martinez-Criado, M. Burghammer, und C. Ronning, „Enhanced sputtering and incorporation of Mn in implanted GaAs and ZnO nanowires", *J. Phys. Appl. Phys.* **2014**, 47 (39), 394003. DOI: 10.1088/0022-3727/47/39/394003

[2]     C. Borschel und C. Ronning, „Ion beam irradiation of nanostructures – A 3D Monte Carlo simulation code", *Nucl. Instrum. Methods Phys. Res. Sect. B Beam Interact. Mater. At.* **2011**, 269 (19), p. 2133–2138. DOI: 10.1016/j.nimb.2011.07.004